\documentclass[pra,
twocolumn,
showpacs,
floatfix]{revtex4-1}
\usepackage[utf8]{inputenc}
\usepackage{color}

\usepackage{lmodern}
\usepackage{amsmath}
\usepackage{amssymb}
\usepackage{mathtools}

\usepackage[english]{babel} 

\usepackage[tight,nice]{units}

\newcommand{\mzh}[1]{\times\nobreak10^{#1}}
\newcommand{\eref}[1]{(\ref{#1})}

\newcommand{\Real}{\mathrm{Re}}

\newcommand{\abs}[1]{\left| #1 \right|}

\DeclareMathAlphabet{\bi}{OML}{cmm}{b}{it}
\renewcommand{\vec}[1]{ \bi{ #1 } }
\newcommand{\rme}{\mathrm{e}}
\newcommand{\rmi}{\mathrm{i}}
\newcommand{\rmd}{\mathrm{d}}

\newcommand{\bnabla}{\boldsymbol{\nabla}}

\newcommand{\imagi}{\rmi}

\renewcommand{\vr}{\vec{r}}

\newcommand{\vv}{\vec{v}}
\newcommand{\ve}{\vec{e}}
\newcommand{\vp}{\vec{p}}

\newcommand{\vk}{\vec{k}}
\newcommand{\vF}{\vec{F}}
\newcommand{\vE}{\vec{E}}
\newcommand{\vA}{\vec{A}}
\newcommand{\vB}{\vec{B}}
\newcommand{\valpha}{\boldsymbol{\alpha}}
\newcommand{\ex}{\bi{e}_x}
\newcommand{\ey}{\bi{e}_y}
\newcommand{\ez}{\bi{e}_z}

\newcommand{\Ip}{I_{\mathrm{p}}}

\newcommand{\half}{\frac{1}{2}}

\newcommand{\beq}{\begin{equation}}
\newcommand{\eeq}{\end{equation}}

\begin{document}

\title{Coulomb-corrected strong-field quantum trajectories beyond dipole approximation}

\author{Th.~Keil}

\author{D.~Bauer}
\affiliation{Institut f\"ur Physik, Universit\"at Rostock, 18051 Rostock, Germany}

\date{\today}

\begin{abstract} 
  
Non-dipole effects in strong-field photoelectron momentum spectra have been revealed experimentally [C.T.L.\ Smeenk {\em et al.}, Phys.\ Rev.\ Lett.\ {\bf 106}, 193002 (2011); A.\ Ludwig {\em et al.}, Phys.\ Rev.\ Lett.\ {\bf 113}, 243001 (2014)]. For certain laser parameters and photoelectron momenta the spectra were found to be shifted {\em against} the laser propagation direction whereas one would naively assume that the radiation pressure due to the $\vv\times\vB$-force pushes electrons always {\em in} propagation direction. Only the interplay between Lorentz and Coulomb force may give rise to such counterintuitive dynamics.
  In this work, we calculate the momentum-dependent shift in and against the propagation direction by extending the quantum trajectory-based Coulomb-corrected strong-field approximation beyond the dipole approximation. A semi-analytical treatment where both magnetic and Coulomb force are treated perturbatively but simultaneously reproduces the results from the full numerical solution of the equations of motion. 
\end{abstract}

\maketitle

\section{Introduction}

The dipole approximation is frequently applied in strong-field physics when the laser wavelength $\lambda$ is much larger than the electron excursion, or the electron velocity is much smaller than the speed of light $c$. Both conditions lead to  $A_0/c \ll 1$ where $A_0$ is the vector potential amplitude. For instance, for $\lambda=3.4\,\mu$m and a laser intensity $\unitfrac[6\mzh{13}]{W}{cm^2}$ (the parameters used below) one has $A_0/c \simeq 0.02$ and one would think that non-dipole effects are hardly observable. Nevertheless recent experiments \cite{Smeenk2011,Ludwig2014} have shown them. As for linear polarization photoelectron momentum spectra for isolated atoms calculated in dipole approximation have azimuthal symmetry about the laser polarization axis, any deviation from that symmetry is a signature of  non-dipole effects. Those could be due to an inhomogeneous electric field, which we do not consider here, or due to the $\vv\times\vB$ force, i.e., the radiation pressure or ponderomotive force \cite{Mulser2010}, in which we are interested in this work. The radiation pressure is expected to break the symmetry with respect to the propagation direction, say, $\ve_y$, i.e., the spectrum is not symmetric anymore under the reflection $p_y \to -p_y$.   It has been found that such asymmetries are caused by the photon momentum partitioning between electron and ion \cite{Titi2012,Liu2013}, or classically speaking, by the momentum transfer due to the magnetic component of the Lorentz force.  A relativistic calculation based on  the time-dependent Dirac equation was carried out in \cite{Ivanov2015}, confirming these findings. A non-dipole strong-field approximation using the exact non-dipole Volkov solution for the Schrödinger equation was developed in \cite{He2017}, predicting the action of the radiation pressure and including Coulomb effects via rescattering. The experimental results presented in \cite{Ludwig2014} show that for certain laser intensities and electron momenta the photoelectron spectra are counterintuitively shifted \emph{against} the laser propagation direction, which is also seen in the results of \cite{Smeenk2011}. More detailed studies using numerical solutions of the time-dependent Schr\"odinger equation (TDSE) and classical trajectories \cite{Chelkowski2015} demonstrate that the so-called ``long trajectories'' (which pass the ion at least once before reaching the detector) may be pushed against the propagation direction while ``short trajectories'' (moving directly to the detector) are pushed in  propagation direction. 

In this article, we extend our previously introduced quantum trajectory-based Coulomb-corrected strong-field approximation (TCSFA)  \cite{Popruzhenko2008,Popruzhenko2008b,Yan2010,Yan2013,Keil2014} to calculate the counterintuitive momentum shifts of photoelectrons. To that end the magnetic $\vv\times\vB$ force is added to the equations of motion (EOM) for the trajectories from the ``tunnel exit'' to the detector. We show that a semi-analytical, perturbative solution of the EOM taking both Coulomb and magnetic force into account is sufficient to reproduce the momentum shifts.

\section{Non-Dipole Quantum Trajectories}
The TCSFA is based on the strong-field approximation (SFA) in saddle-point approximation (see, e.g., \cite{Milosevic2006,Popruzhenko2014a}) where the photoelectron spectrum is calculated as a sum over saddle-points, each of which can be associated with a quantum trajectory in complex space and time. In plain SFA, the Coulomb potential is neglected after ionization has occurred at complex time $t_s$ according to the saddle-point equation \eqref{eq:spe}, discussed below.  Within the TCSFA, the trajectories are corrected by  propagation according to Newton's EOM, including the Coulomb force. The individual contributions from each saddle point $s$ are calculated as
\begin{align}
  \label{eq:11}
  M_\vp(t_s) &= f_{\Psi_0} \rme^{-\imagi S_{\vp,\Ip}(t_s)/\hbar}.
\end{align}
Here, $f_{\Psi_0}$ is a pre-exponential factor depending on the initial electronic state $\Psi_0$  (see, e.g., \cite{Milosevic2006}), and $S_{\vp,\Ip}(t)$ is the action integral defined in Eq.~\eref{eq:tcsfa:action} below. The photoelectron momentum spectrum then is the modulus square of the coherent sum of these contributions,
 \begin{align}
  \label{eq:totspec}
  |M_\vp|^2 &= \left| \sum_s   M_\vp(t_s) \right|^2,
\end{align}
allowing for interference effects.

Our procedure to include non-dipole effects consists of two parts. The first part is to choose a position-dependent vector potential that describes a linearly polarized laser pulse, for instance
\begin{align}
  \label{eq:1}
  \vA(\vr,t) &= A_0 \ez \sin^2(\Phi)\sin(\phi)
\end{align}
for $0<\Phi<\pi$ and zero otherwise, with 
\begin{align}
\Phi=\Omega t - K y, \quad \phi=\omega t -ky + \varphi.
\end{align}
 The polarization direction is $\ez$, and the propagation direction  is  $\ey$. The envelope frequency reads $\Omega = \omega / (2n_c)$ with $n_c$ the number of laser cycles in the pulse, the envelope wave vector is accordingly  $\vec{K}=\vk/(2n_c)$ with  $\vk = \ey\omega/c$. The corresponding electric and magnetic fields read
\begin{align}
  \label{eq:2}
  \vE&(\vr,t) = -\frac{\partial}{\partial t} \vA(\vr,t) \\
             &= -A_0\omega \ez \sin(\Phi) \left(\frac{1}{n_c}\cos(\Phi)\sin(\phi) + \sin(\Phi)\cos(\phi) \right)\nonumber\\
  \vB&(\vr,t) = \bnabla \times \vA(\vr,t) = \frac{1}{c} ( \vE(\vr,t)\cdot\ez)\ex .\nonumber
\end{align}
The carrier-envelope phase $\varphi$ is set to $0$ in the following. 

The second part in our procedure is the modification of the trajectory propagation. The EOM 
\begin{align}
  \label{eq:3}
  \frac{\rmd}{\rmd t} \vr &= \vv\,,\quad \frac{\rmd}{\rmd t} \vp = \vF
\end{align}
keep their form but the relation between the velocity $\vv$ and the kinetic momentum $\vp$ reads 
\begin{align}
  \label{eq:4}
  \vp &= \gamma m_\rme \vv
\end{align}
now, 
where $m_\rme$ is the rest mass of the electron, and
\begin{align}
  \label{eq:5}
  \gamma &= \frac{1}{\sqrt{1-\frac{v^2}{c^2}}}= \sqrt{1+\frac{p^2}{m_\rme^2c^2}}
\end{align}
is the Lorentz factor. The energy reads 
\begin{align}
  \label{eq:6}
  E & = \gamma m_\rme c^2 = \sqrt{p^2c^2 + m_\rme^2c^4},
\end{align}
and the EOM
\begin{align}
  \label{eq:8}
    \frac{\rmd}{\rmd t} \vr &= \frac{\vp}{m_\rme\sqrt{1+\frac{p^2}{m_\rme^2c^2}}}\\
  \frac{\rmd}{\rmd t} \vp &= - e\left( \vE + \frac{\vp}{m_\rme\sqrt{1+\frac{p^2}{m_\rme^2c^2}}} \times \vB \right) - \bnabla V(\vr)
\end{align}
where $e$ is the absolute value of the electron charge and
\begin{align}
  \label{eq:12}
  V(\vr) &= - \frac{eZ}{\sqrt{\vr^2(t)}}
\end{align}
is the Coulomb force (retardation effects being neglected). The magnetic $\vv \times\vB$ term in the Lorentz force $\vF_\mathrm{L}=-e(\vE+\vv \times\vB)$ is of order $O(\frac{v}{c})$ because $\abs{\vB}=\frac{1}{c}\abs{\vE}$. True relativistic effects are of order $O(\frac{v^2}{c^2})$, for instance in the kinetic energy
\begin{align}
E - m_\rme c^2 &= \frac{\vp^2}{2m_\rme} + O\left(\frac{v^2}{c^2}\right).
\end{align}
We restrict ourselves to non-dipole effects of order $O(\frac{v}{c})$ as the plain SFA underlying the TCSFA is based on the TDSE and not the Dirac equation. An extension into the relativistic regime should be possible though since  a relativistic SFA using the original Volkov solutions \cite{Wolkow35} has been developed early on \cite{Reiss1990}.
 
The quantum trajectories in the saddle-point SFA and TCSFA carry a phase, allowing them to interfere when added up coherently according \eqref{eq:11}. In the TCSFA this phase is the Coulomb-corrected plain-SFA action  \cite{Popruzhenko2008,Popruzhenko2008b,Yan2010,Yan2013,Keil2014}
\begin{align}
  \label{eq:tcsfa:action}
  S_{\vp,\Ip}(t_s) &= \int_{t_s}^\infty \rmd t\,\left( \frac{\vp^2(t)}{2m_\rme} + V(\vr) + \Ip \right)
\end{align}
where $\Ip$ is the ionization potential and, as stated above, relativistic corrections of order $O(\frac{v^2}{c^2})$ are neglected.  

The saddle-point (or ionization) times $t_s$ are determined by the plain-SFA saddle-point equation  \cite{Milosevic2006,Popruzhenko2014a}
\begin{align}
  \label{eq:spe}
  [\vp_{\mathrm{drift}}+e\vA(t_s)]^2 &= -2\Ip
\end{align}
where the position dependence of the vector potential  has been dropped, i.e., $\vA(t_s)=\vA(y=0,t_s)$. As $\vp_{\mathrm{drift}}$ is real and $\Ip>0$, the solutions $t_s$ are complex. Details are discussed in, e.g., \cite{Popruzhenko2008,Yan2010}. The propagation may be split into a sub-barrier part from $t_s$ to $\Real\, t_s$ (which may be identified with the tunneling step) and the motion from the ``tunnel exit'' at $\Real\, t_s$ to the detector at $t\rightarrow\infty\,$. We restrict ourselves to corrections to the latter part of the quantum trajectories, which turns out to be sufficient to understand the  observed non-dipole drifts in and against the propagation direction of the laser pulse on a qualitative level. The initial conditions for the real time propagation from $\Real\,  t_s$ to $T$ with $\vA(\vr,T)=0$ are chosen 
\begin{align}
  \label{eq:13}
  \vp(\Real\, t_s) &= \vp_{\mathrm{drift}}+e\vA(\Real\, t_s),\\
  \vr(\Real\, t_s) &= \valpha(\Real\, t_s) - \Real\, \valpha(t_s) \label{eq:13b}
\end{align}
with the elongation 
\begin{align}
  \label{eq:14}
  \valpha(t) &= \frac{e}{m_\rme}\int^t \rmd t' \vA(t')\,.
\end{align}
Equation \eqref{eq:13b} ensures that time and position are purely real at the tunnel exit; we note in passing though that the choice for the initial condition of $\vr(t)$ should be different if the sub-barrier part of the ionization dynamics needs also to be Coulomb-corrected \cite{Popruzhenko2008,Torlina2012,Pisanty2016}, as it is the case for the derivation of quantitatively correct ionization rates employing matching to the ground state wavefunction \cite{Popov2004,Popruzhenko2008a}. Further, within the TCSFA the problem of branch cuts \cite{Popruzhenko2014,Pisanty2016,Keil2016} in the complex-time plane due to the square-root in the denominator  in \eqref{eq:12} is avoided by considering only the Coulomb correction to the purely real trajectories \eqref{eq:13b} beyond the tunnel exit.  
  
Whereas in dipole approximation it is sufficient to set $T$ to the pulse duration $T_{\mathrm{p}}=2\pi/\Omega$, here one needs to ensure that for all propagated trajectories the laser pulse has passed $\vr(T)\,$. However, since the laser pulse travels with $c$ but the final velocities of the electrons are much smaller, it is more than sufficient to choose, e.g., $T=2T_{\mathrm{p}}\,$. As we ignore relativistic $O(\frac{v^2}{c^2})$ effects, once $\vA=0$ the further propagation in the Coulomb potential alone can be carried out analytically using Kepler's laws.

For brevity atomic units ($\hbar=m_\rme=e=4\pi\varepsilon_0=1$) are used from here on unless noted otherwise.

\section{Spectra}
\begin{figure}[h]
  \centering
  \includegraphics[scale=0.75]{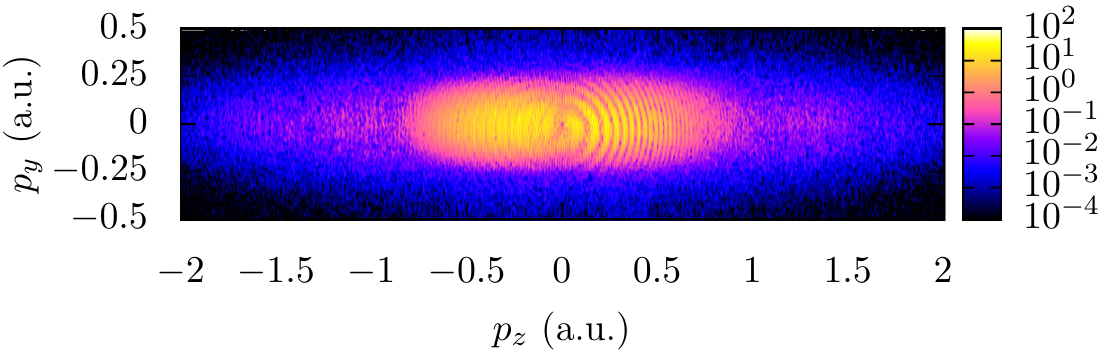}
  \caption{Momentum-resolved photoelectron spectrum $|M_\vp|^2$, Eq.~\eqref{eq:totspec}, for xenon ($\Ip=0.447$) irradiated with a six-cycle $\sin^2$-shaped laser pulse of  intensity ${I=\unitfrac[6\mzh{13}]{W}{cm^2}}\,$, wavelength $\lambda=\unit[3.4]{\mu m}\,$. The momentum $p_z$ is in polarization direction, $p_y$ in propagation direction of the laser pulse.}
  \label{fig:spectra:numerical}
\end{figure}

Using the method just introduced, we calculated photoelectron spectra for some of the parameters of Ref.~\cite{Ludwig2014}. The target is a xenon atom with ionization potential $\Ip=0.447$ being irradiated by a six-cycle $\sin^2$-shaped laser pulse with an intensity of $I=\unitfrac[6\mzh{13}]{W}{cm^2}$ and a wavelength of $\lambda=\unit[3.4]{\mu m}$. $N=6\mzh{7}$ trajectories were propagated to generate the momentum-resolved spectrum shown in Fig.~\ref{fig:spectra:numerical}. 

\begin{figure}[h]
  \centering
  \includegraphics[scale=0.75]{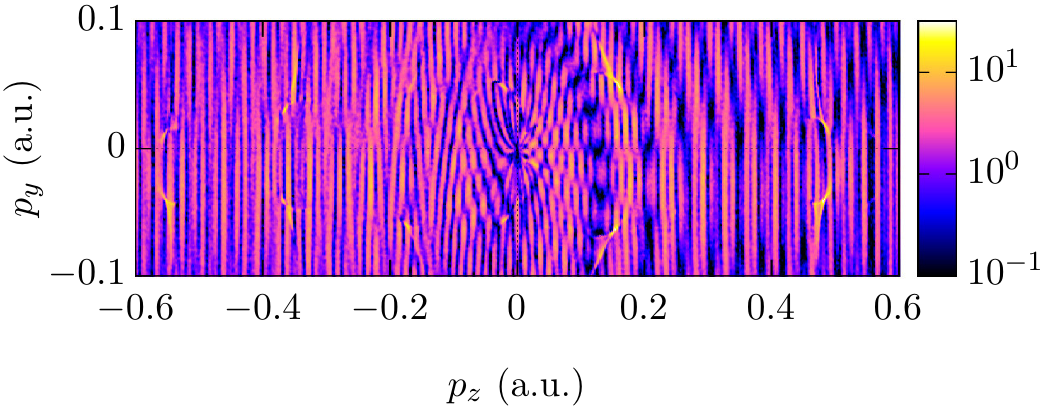}
  \caption{Same as Fig.~\ref{fig:spectra:numerical} but for a smaller momentum range and calculated with higher resolution.}
  \label{fig:spectra:numerical:zoom}
\end{figure}

Since the non-dipole effects are expectedly small, we show a close-up view of the central part of Fig.~\ref{fig:spectra:numerical} in Fig.~\ref{fig:spectra:numerical:zoom}.

Asymmetries under reflection in polarization direction $p_z \to -p_z$ arise already in dipole approximation for short pulses whereas there is strict azimuthal symmetry about the polarization axis.  Instead, Fig.~\ref{fig:spectra:numerical:zoom} reveals a clear asymmetry with respect to the propagation direction $\ey$, which is a clear non-dipole effect. The shift is easily notable, for example, from the semiclassical caustics appearing at $p_{z}\simeq -0.55$ and $p_{z}\simeq 0.5$, the so-called  low-energy structures (LES) \cite{Kaestner2012,Moeller2014,Zhang2016}. These are significantly shifted against propagation direction.

\begin{figure}[h]
  \centering
  \includegraphics[scale=0.75]{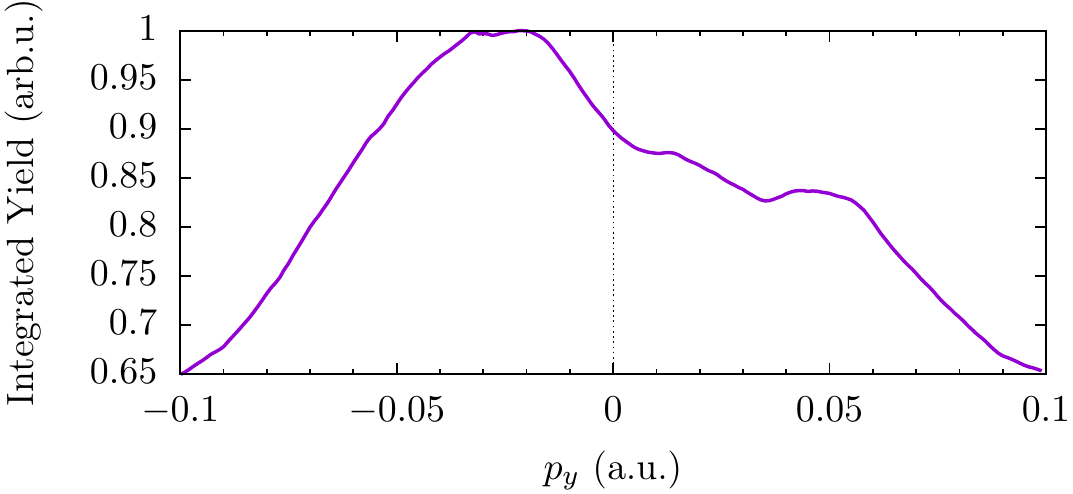}
  \caption{Same parameters as in Fig.~\ref{fig:spectra:numerical} and \ref{fig:spectra:numerical:zoom} but integrated over $p_z$. }
  \label{fig:spectra:numerical:integrated}
\end{figure}

To visualize the total offset of the spectrum it is useful to integrate over $p_{z}$ which yields a projection onto the $p_{y}$-axis \cite{Ludwig2014}. This is shown in Fig.~\ref{fig:spectra:numerical:integrated}.
We clearly observe a small shift of the integrated spectrum in negative propagation direction. The magnitude of that shift is on the same order as the experimental one shown in \cite{Ludwig2014}.

\begin{figure}[h]
  \centering
  \includegraphics[scale=0.75]{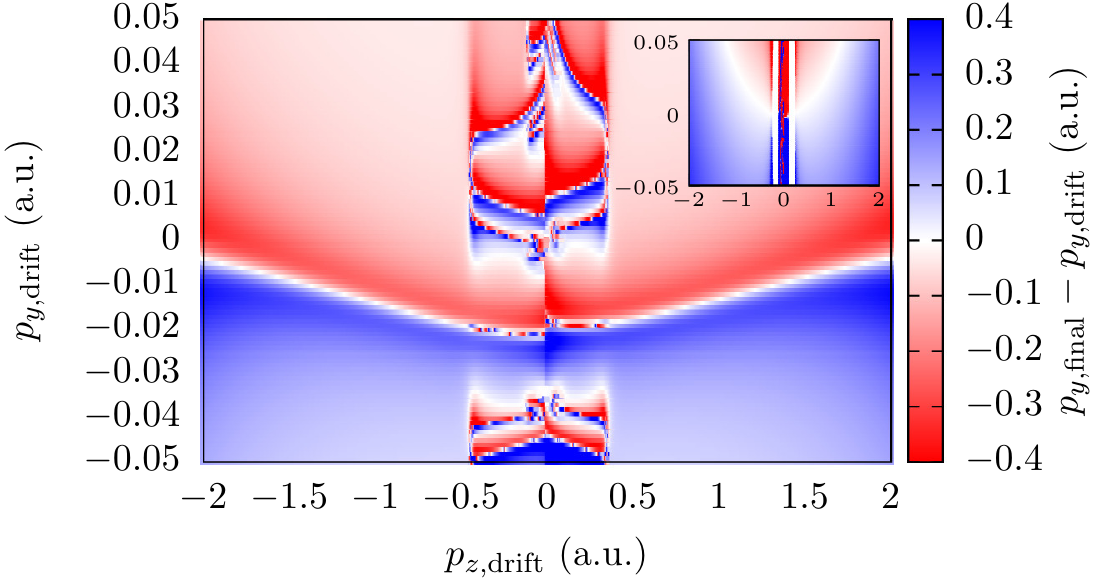}
  \caption{Difference between initial drift momentum (equal to the final momentum in plain SFA) and final momentum, plotted in the plane of the initial drift momentum. Blue color indicates a shift {\em in} propagation direction, red {\em against} propagation direction. Chaotic behavior is seen in the central part which can be attributed to multiple returns happening only for small drift momenta. The main panel refers to long trajectories, the inset to short trajectories. The color scale for the inset spans only one twentieth of that in the main panel. The missing parts (white vertical lines) for small $p_{z,\mathrm{drift}}$ correspond to bound trajectories which never reach the detector.}
  \label{fig:spectra:shift:numerical}
\end{figure}

The shift of the spectrum can be understood by analyzing the difference between the corrected and the corresponding plain-SFA trajectories. In \cite{Liu2013,Chelkowski2015} it was shown that the short trajectories are shifted {\em in} positive propagation direction, as intuitively expected from radiation pressure, whereas long trajectories are shifted \emph{against} propagation direction. This behavior is shown in detail in Fig.~\ref{fig:spectra:shift:numerical} where the shift $p_{y,\mathrm{final}} - p_{y,\mathrm{drift}}$ is plotted vs the initial drift momentum.
In the plain SFA the final momentum $\vp_{\mathrm{final}}$ is equal to the initial drift momentum $\vp_{\mathrm{drift}}$ so that we can infer directly from Fig.~\ref{fig:spectra:shift:numerical} for each plain-SFA trajectory in which direction it is shifted due to Coulomb and non-dipole effects.  The main panel in Fig.~\ref{fig:spectra:shift:numerical}  shows shifts for the long trajectories. The central region for $\abs{p_{z,\mathrm{drift}}}<0.5$ is dominated by chaotic behavior. This can be attributed to trajectories exhibiting multiple interactions with the ion, which are not of interest here. On the axis $p_{y,\mathrm{drift}}=0$ the shift is negative  (neglecting the central, chaotic region) for the  $p_{z,\mathrm{drift}}$ range shown. Only for negative $p_{y,\mathrm{drift}}$ positive shifts are visible. For larger $|p_{z,\mathrm{drift}}|$ the shift is also positive (not shown), as observed in the experiment. The demarcation line of zero shift is displaced to negative drift momenta $p_{y,\mathrm{drift}}$, the magnitude of the displacement is on the same order as the shift of the total spectrum. The inset shows only the short trajectories. These mainly exhibit a positive shift which increases with larger momentum $|p_{z,\mathrm{drift}}|$ in polarization direction. The negative shift in the upper central region is due to the Coulomb attraction, which reduces the total momentum. Note that the color scale for the inset is only one twentieth of that for the main panel so that compared to the long trajectories the shift of the short trajectories is negligible. Accordingly we focus in the following on the long trajectories and whether their behavior can be described analytically.

\section{SFA-based Analytical Model}
In the above simulations both magnetic-field and Coulomb effects are incorporated. Neither Coulomb effects alone nor non-dipole effects alone can explain the counterintuitive shift against the laser propagation direction. We build our analytical model on the assumption that both Coulomb and magnetic forces on the unperturbed SFA trajectory are perturbative. The   unperturbed SFA trajectory  in dipole approximation (indicated by a subscript $0$) is given by
\begin{align}
  \label{eq:sfatrajectory}
  \vv_0(t) &= \vp_{\mathrm{drift}}+\vA(t),\\
  \vr_0(t) &= \vp_{\mathrm{drift}}\,(t-t_s) + \int_{t_s}^t \rmd t'\,\vA(t') + \vr_{\mathrm{initial}}.
\end{align}
Using the same initial conditions at time $\Real\, t_s$ as above we can write the total momentum shift due to the magnetic component of the Lorentz force in lowest order as
\begin{align}
  \Delta p_y^{\mathrm{L}}(t) &= -\int_{\Real\,  t_s}^t \rmd t'\, \left[ \vv_0(t') \times \vB(t') \right]\cdot\ey
\end{align}
where $\vB(t)=\vB(y=0,t)$. Using $\vB(t)=\ex\frac{1}{c}E_z(t)\,$, $\vE(t)=-\partial \vA(t)/\partial t$, and the unperturbed trajectory $\vv_0(t)$ from \eref{eq:sfatrajectory} this yields
\begin{align}
  \Delta p_y^{\mathrm{L}}(t) &= \frac{1}{c} \int_{\Real\,  t_s}^t \rmd t'\, \left(p_{z,\mathrm{drift}} + A_z(t')\right)\left(\frac{\rmd}{\rmd t'} A_z(t') \right)\nonumber \\
 &= \frac{1}{c}\left[ p_{z,\mathrm{drift}} A_z(t') + \half A_z^2(t') \right]_{\Real\,  t_s}^t
\end{align}
with $ A_z(t)= A_z(y=0,t) $.
For $t\rightarrow\infty$ both terms vanish due to $\vA(t\rightarrow\infty)=0$ so that the total shift $\Delta p_{y,\mathrm{total}}^{\mathrm{L}} = \Delta p_y^{\mathrm{L}}(t\rightarrow\infty)$ reads
\begin{align}
  \label{eq:model:lorentztotalmomentumshift}
  \Delta p_{y,\mathrm{total}}^{\mathrm{L}} &= -\frac{1}{c}\left[ p_{z,\mathrm{drift}} A_z(\Real\,  t_s) + \half A_z^2(\Real\,  t_s) \right]\,.
\end{align}

Having calculated the momentum shift due to the magnetic field, we can include approximately the influence of the Coulomb potential. The main difference between short and long trajectories in the plain SFA is that long trajectories pass the ion (at least) once before reaching the detector whereas short trajectories do not. It is  this flyby that causes the differences in the shifts as seen in Fig.~\ref{fig:spectra:shift:numerical}. Assuming a sufficiently fast flyby, the corresponding trajectory $\vr_f(t)$ can be approximated as
\begin{align}
  \label{eq:15}
  \vr_f(t) &= y_f\ey + p_{z,f}(t-t_f)\ez
\end{align}
where $t_f$ is the flyby time defined by 
\begin{align}
  \label{eq:flybytime}
  \vr_0(t_f)\cdot\ez=0\,, 
\end{align}
$y_f$ is the distance in propagation direction at flyby and $p_{z,f}=\nobreak\vp_0(t_f)\cdot\ez$ is the respective momentum in polarization direction. To evaluate the momentum shift we integrate over the Coulomb force $\vF^{\mathrm{C}}(\vr)=-\bnabla V(\vr)$ along the flyby trajectory $\vr_f(t)$,
\begin{align}
  \label{eq:flybyintegral}
  \Delta p_y^{\mathrm{C},\mathrm{flyby}} &= -\int_{t_f-\Delta t}^{t_f+\Delta t} \rmd t\,\frac{y_f}{\left( y_f^2 + p_{z,f}^2(t-t_f)^2 \right)^{3/2}}\,.
\end{align}
This integral can be solved analytically,
\begin{align}
  \label{eq:17}
  \Delta p_y^{\mathrm{C},\mathrm{flyby}} &= \left. - \frac{t-t_f}{y_f\sqrt{y_f^2+p_{z,f}^2(t-t_f)^2} }\right|_{t_f-\Delta t}^{t_f+\Delta t}\nonumber \\
&= -\frac{2\Delta t}{y_f\sqrt{y_f^2+p_{z,f}^2\Delta t^2}} \,.
\end{align}
For $p_{z,f}^2\Delta t^2 \gg y_f^2 $  we obtain the simple result
\begin{align}
  \label{eq:18}
  \Delta p_y^{\mathrm{C},\mathrm{flyby}} &= -\frac{2}{y_f\sqrt{p_{z,f}^2}} \,.
\end{align}
One contribution to $y_f$ is $\vr_0(t_f)\cdot\ey$ from the unperturbed trajectory. However, to account for both Lorentz and Coulomb force we need to add the spatial shift induced by $\Delta p_y^{\mathrm{L}}(t)$ so that $y_f=\vr_0(t_f)\cdot\ey+\Delta y^{\mathrm{L}}(t_f)$ with
\begin{align}
  \label{eq:19}
  \Delta &y^{\mathrm{L}}(t) = \int_{\Real\,  t_s}^t \rmd t'\,\Delta p_y^{\mathrm{L}}(t')\\
  &= \left[\frac{p_{z,\mathrm{drift}}}{c}\alpha(t') + \frac{1}{2c}\alpha^{(2)}(t') + \Delta p_{y,\mathrm{total}}^{\mathrm{L}}\,t'\right]_{\Real\,  t_s}^t
\end{align}
where we defined
\begin{align}
  \label{eq:20}
  \alpha(t) = \int^t\rmd t'\,A_z(t')\,,\quad\alpha^{(2)}(t) = \int^t\rmd t'\,A_z^2(t')\,.
\end{align}
For the $\sin^2$-shaped laser pulse used above this can be evaluated analytically. Solely the flyby time $t_f$ needs to be calculated numerically due to the transcendental nature of \eref{eq:flybytime}.

The reduction of the total momentum due to the long-range Coulomb potential can be approximated by a simple estimate. Assuming that only the kinetic energy is affected without any influence on the direction we can write the final kinetic energy as the initial kinetic (drift) energy plus the (negative) potential energy at the tunnel exit, which yields
\begin{align}
  \abs{\vp_{\mathrm{fin}} } &= \sqrt{\vp_{\mathrm{drift}}^2 + 2V(\vr(\Real\,  t_s))}\,.
\end{align}
This amounts to a change in the momentum in propagation direction of
\begin{align}
  \Delta p_y^{\mathrm{C},\text{long-range}} &= \left(\frac{\sqrt{\vp_{\mathrm{drift}}^2 + 2V(\vr(\Real\,  t_s)) }}{\sqrt{\vp_{\mathrm{drift}}^2}} - 1 \right)p_{y,\mathrm{drift}}\,.
\end{align}

Using the equations derived above we can calculate the momentum shift in propagation direction predicted from our simple model as
\begin{align}
  \label{eq:model:totalmomentumshift}
  \Delta p_y^{\mathrm{total}} &= \Delta p_{y,\mathrm{total}}^{\mathrm{L}} + \Delta p_y^{\mathrm{C},\mathrm{flyby}} +\Delta p_y^{\mathrm{C},\text{long-range}}\,.
\end{align}

\begin{figure}[h]
  \centering
  \vspace{2mm}
  \includegraphics[scale=0.75]{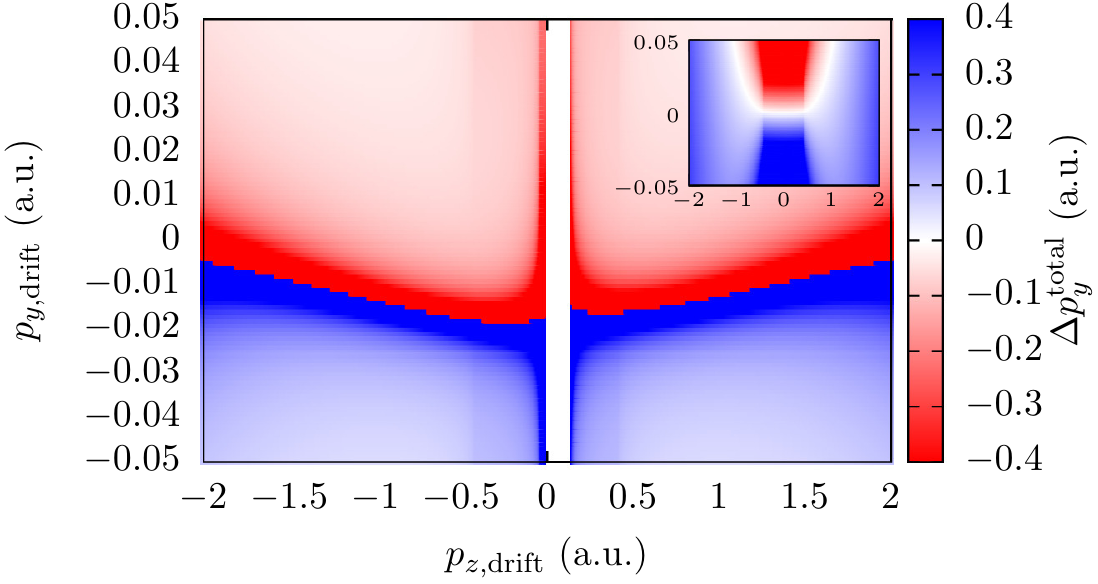}
  \caption{Same as Fig.~\ref{fig:spectra:shift:numerical}, but calculated from Eq.~\eref{eq:model:totalmomentumshift} instead of a full numerical solution of the EOM. The main panel shows the shift for long trajectories (assuming one flyby).  Trajectories with too small positive $p_{z,\mathrm{drift}}$  do not complete their flyby during the pulse and are not considered (white area).  The inset shows the momentum shift $\Delta p_{y,\mathrm{total}}^{\mathrm{L}}$ from \eref{eq:model:lorentztotalmomentumshift} for short trajectories (no flyby). The color scale for the inset spans only one twentieth of that in the main panel.}
  \label{fig:spectra:shift:analytical}
\end{figure}

For the numerical evaluation we use the same momentum grid as in Fig.~\ref{fig:spectra:shift:numerical} and calculate the most probable ionization time $t_s$ corresponding to a long trajectory for every momentum $\vp_{\mathrm{drift}}$ on the grid from the saddle-point equation \eref{eq:spe}. Then we find the first flyby time $t_f$ from \eqref{eq:flybytime} and evaluate the respective shift $\Delta p_y^{\mathrm{total}}$. The result is shown in Fig.~\ref{fig:spectra:shift:analytical}. 

 For sufficiently high momenta in polarization direction the demarcation line in Fig.~\ref{fig:spectra:shift:numerical} between positive and negative shifts is reproduced very well by our perturbative, semi-analytical model.
The inset shows the momentum shift for short trajectories where $\Delta p_y^{\mathrm{C},\mathrm{flyby}}=0\,$. The estimate for $\Delta p_y^{\mathrm{C},\text{long-range}}$ allows to qualitatively reproduce the structure seen in the inset of Fig.~\ref{fig:spectra:shift:numerical}. Without that term only $\Delta p_{y,\mathrm{total}}^{\mathrm{L}}$ remains, which shows the same behavior for all $p_{y,\mathrm{drift}}$, i.e., the same as for $p_{y,\mathrm{drift}}=0\,$. Instead, for the long trajectories the term $\Delta p_y^{\mathrm{C},\text{long-range}}$ is negligible.

\section{Conclusions}
We extended our trajectory-based  Coulomb-corrected strong-field approximation for the inclusion of non-dipole effects. In particular, we analyzed the counterintuitive shift of final photoelectron momenta against propagation direction in terms of quantum trajectories originating from a non-dipole and Coulomb-corrected strong-field approximation. The full numerical solution of the equations of motion including both magnetic and Coulomb force from the tunnel exit on reproduces the experimentally observed shifts in and against the propagation direction of the laser pulse, depending on the photoelectron momentum. These shifts are also obtained within a simplified semi-analytical model where first the shift in propagation direction due to the magnetic field is calculated perturbatively, followed by a perturbative Coulomb correction to the long trajectories when they encounter the parent ion during flyby. In this way, we confirm  that the mechanism which shifts the photoelectron spectra against propagation direction is the combination of radiation pressure and a subsequent soft recollision. Our results show that non-dipole effects can be easily incorporated into trajectory-based methods, which---complementary to the brute-force numerical solution of the time-dependent Schr\"odinger equation---allow for intuitive interpretations and thus lead to better understanding of the underlying physical mechanisms.

\section*{Acknowledgment}
The authors acknowledge support through the SFB~652 and project BA~2190/8 of the German Science Foundation (DFG).

\bibliography{bibliography}

\end{document}